\documentclass[10pt,a4paper]{article}

\usepackage[a4paper,margin=2cm]{geometry}
\usepackage{amsmath,amssymb}
\usepackage{graphicx}

\usepackage[utf8x]{inputenc}

\usepackage{hyperref}

\usepackage{array}

\newcolumntype{+}{!{\vrule width 2pt}}

\newlength\savedwidth

\usepackage{hyperref}
\providecommand{\href}[2]{\texttt{#2}}
\providecommand{\url}[1]{\texttt{#1}}

\usepackage{float}

\graphicspath{{images_new/}{../images_new/}}

\usepackage[labelfont=bf,labelsep=period]{caption}   
\usepackage{subcaption}

\usepackage{multirow}

\usepackage[official]{eurosym}

\newcommand{\bea}{\begin{eqnarray}}
\newcommand{\eea}{\end{eqnarray}}
\newcommand{\beq}{\begin{equation}}
\newcommand{\eeq}{\end{equation}}

\providecommand{\eqref}[1]{(\ref{#1})}
\newcommand{\figref}[1]{Fig \ref{#1}}
\newcommand{\Figref}[1]{Fig \ref{#1}} 
\newcommand{\tabref}[1]{Table~\ref{#1}}
\newcommand{\secref}[1]{Section~\ref{#1}}

\newcommand{\tnote}[1]{} 
\newcommand{\tcomment}[1]{} 
\newcommand{\qnote}[1]{} 

\newcommand{\paperdefn}[1]{\textsc{#1}}

\newcommand{\zbar}{\bar{z}}
\newcommand{\ellmax}{\ell_\mathrm{max}}
\newcommand{\NLCC}{N_\mathrm{LCC}}
\newcommand{\texpect}[1]{\langle #1 \rangle}

\newcommand{\notationmat}[1]{{\mathbf{\textsf{#1}}}}
\newcommand{\Pmat}{\notationmat{P}}

\newcommand{\notationset}[1]{{\mathcal{#1}}}
\newcommand{\Bcal}{\notationset{B}}
\newcommand{\Ccal}{\notationset{C}}

\newcommand{\Pcal}{\notationset{P}}

\newcommand{\Scal}{\notationset{S}}
\newcommand{\Tcal}{\notationset{T}}
\newcommand{\Vcal}{\notationset{V}}

\begin{document}
\vspace*{0.2in}

\begin{flushleft}
{\Large
\textbf\newline{How the network properties of shareholders vary with investor type and country} 
}
\newline
\\
Qing Yao\textsuperscript{1,2,*},
Tim S.\ Evans\textsuperscript{1,2},
Kim Christensen\textsuperscript{1,2}
\\
\bigskip
\textbf{1} \href{https://www.imperial.ac.uk/complexity-science}{Centre for Complexity Science}, Imperial College London, London, SW7 2AZ, U.K.
\\
\textbf{2} \href{https://www.imperial.ac.uk/condensed-matter-theory}{Blackett Laboratory}, Imperial College London, London, SW7 2AZ, U.K.
\\
\bigskip

%
%





* q.yao15{@}imperial.ac.uk

\end{flushleft}
\section*{Abstract}
We construct two examples of shareholder networks in which shareholders are connected if they have shares in the same company.  We do this for the shareholders in Turkish companies and we compare this against the network formed from the shareholdings in Dutch companies. We analyse the properties of these two networks in terms of the different types of shareholder.
We create a suitable randomised version of these networks to enable us to find significant features in our networks.  For that we find the roles played by different types of shareholder in these networks, and also show how these roles differ in the two countries we study.



\section*{Introduction}\label{sec:intro}

Complex networks capture information about the bilateral relations between pairs of objects \cite{BRMW13}. As pairwise relationship are so fundamental to many processes, the networks approach has proved to be a powerful tool for many different areas, see for instance  Newman~\cite{newman2010networks} for an overview.

This paper looks at some networks in an economics context which is one area where networks have proved useful \cite{arthur1999complexity,farmer2012complex,acemoglu2016networks}.
In our work we focus on the networks representing the interactions between companies, a topic that has already received some attention. Vitali, Glattfelder and Battiston used network science to show that the world is in control of a few important shareholders \cite{vitali2011network,glattfelder2009backbone}. Takayasu and her collaborators have  studied the flow of money from suppliers to consumers over long time periods \cite{ohnishi2010network,ohnishi2009hubs,iinoa2010community}.
Viegas \textit{et al}. successfully applied the complex systems theory to Mergers and Acquisitions markets (M\&A), studying the scaling relationship between the companies ancestry and the number of M\&A to predict mergers~\cite{viegas2014dynamics}; Huajiao Li, Pengli An, Haizhong An and \textit{et al}. has studied the common shareholdings and give implications particularly using Chinese listed energy companies~\cite{li2016evolutionary,li2016holding,an2017evolution,li2014shareholding,guan2017information}.

In our work, we use complex network methods to study the investment characteristics of different types of shareholders. To do this we build a network of shareholders linked if they have invested in the same company. The topological structures of this network have been quantified and analyzed.
Furthermore, to provide some insights of organization choices: we have compared measures of the complex network with some empirical analysis.

Before that, we will first summarize some relevant concepts from finance and economics in order to place our work in this context.

\subsection*{The Ownership and Control of Companies}
\label{subsec:owner}

The work of Berle and Means \cite{berle1991modern} provided an early and influential view of how ownership and control of companies need to be linked, based on their perceived failures in 1930's US corporate governance to give shareholders effective control of companies. The approach to company ownership has evolved since the 30's. For instance La Porta et al.\ \cite{porta1999corporate} used information on large corporations in 27 wealthy economies to identify the shareholders with ultimate control of these firms. They came to the conclusion that large shareholders do now typically have power over firms.



%

However, in 1974 Zeitlin's ``Corporate ownership and Control'' \cite{zeitlin1982corporate} proved to be highly influential, undermining previous widely held views that large corporations were not influenced by the wider economic and social environment \cite{mizruchi1992intercorporate}.  A network approach is an ideal way to look at ownership and control within the context of the links between companies and their shareholders. 
A network is made up of two parts: the nodes (vertices or actors), and the bilateral relationship between pairs of nodes which are represented as pairs of nodes known as  edges (links or ties). Nodes can be any appropriate unit while edges can represent any type of relationship between any the units; for example, kinship, material transactions, flow of resources or support. In our context, a natural way to capture the information about shareholdings is to make use the nodes to represent the companies and shareholders. The edges are directed with an edge from a shareholder to the company invested in.

Several studies of shareholders and companies start from this network perspective.  For instance, Vitali, Glattfelder and Battiston~\cite{vitali2011network} use a version of this shareholder-company weighted directed network to capture both the influence by a shareholder through direct shareholdings along with the influence implied by chains of ownership and shareholdings. This investigation first confirms that ownership tends to be parsed among numerous shareholders, while control is found to be in the hands of few important shareholders globally. They discovered a structure of bow-tie, revealing the control flowing to small tightly connected financial institutions.

Another use of this network approach is to study how power may be concentrated in a few financial institutions. Gai and Kapadia~\cite{gai2010contagion} used networks to see if such concentrations led to intrinsic weaknesses in the network of financial institutions which leads to the spread of failure in times of financial stress, using this as a possible explanation of the financial crisis.

\subsection*{Outline of Paper}
\label{subsec:outline}

In this paper, we will use network methods to look at relationship between the shareholders of companies. Our focus will be on companies of all sizes, working with two examples --- two countries with very different economic environments. The economic ecosystem depends on the many smaller companies as much as the few large companies and we include all of these in our data. In the following section, we will discus the source and nature of our data, and how we capture this wider economic environment using a network representation. We will then investigate how network analysis can throw light on the structure of the connections between shareholders.

\section*{Materials and Methods}\label{sec:methods}

\subsection*{Data Sources}\label{subsec:data}

The data used in this research is extracted from the Amadeus, a product of Bureau Van Dijk (BvD)~\cite{BvD}. This provides data on around 21 million companies across Europe, including the names of shareholders, the percentage of a company's shares held by each shareholder.  We have focused on the data for one year, 2014, and two exemplary countries within this data: Turkey and the Netherlands. We found fifty thousand Turkish companies and over a million Dutch companies.

The data we use for macroeconomic statistics is retrieved from the World Bank~\cite{worldbank} and CEIC Data~\cite{CEIC}.

\subsection*{Network Representation}\label{subsec:def}

\begin{figure}[htb]
\centering
\includegraphics[scale=1]{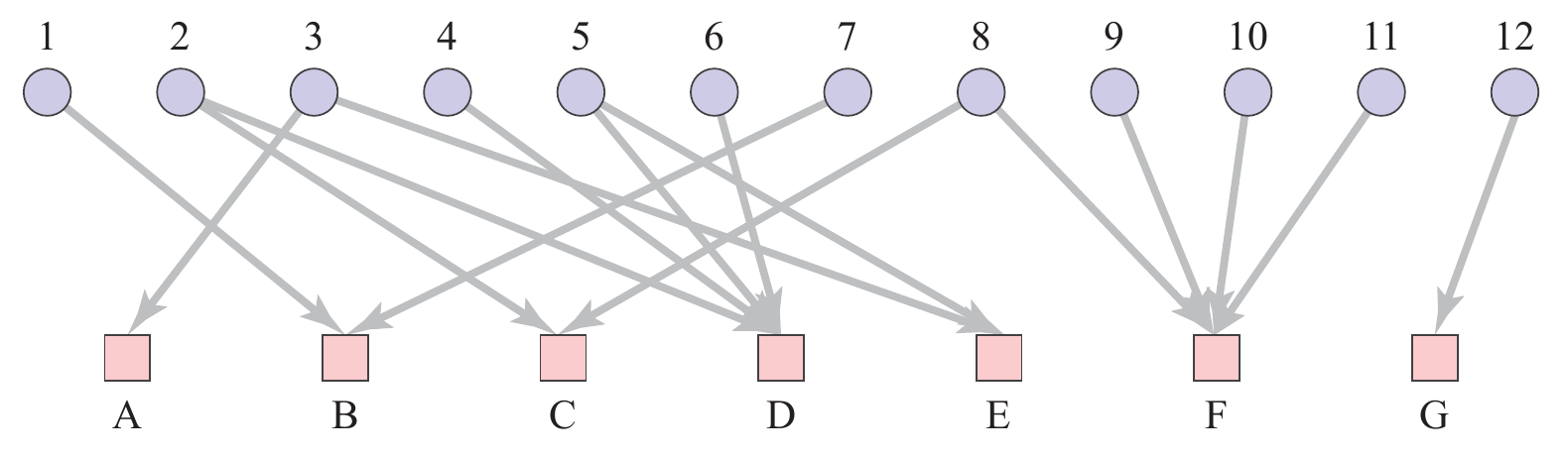}
\caption{{\bf Shareholder-company network example.} An example of our shareholder-company network, which illustrates the relationships between shareholders, the upper blue circles numbered 1 to 12 and companies, the lower red squares labelled A to G.  An edge indicates that there is an investment from the shareholder represented by the source node in the company represented by the target node. For example, shareholders 3 and 5 have both invested in company E, and an arrow represents this relation. In addition, shareholder 3 has invested in company A, but no other shareholder.}
\label{fig:bipartite}
\end{figure}

Our data on the shareholders in companies has a natural representation as a network and this is available at \cite{Y19}.  In our ``shareholder-company  network'', each distinct shareholder and each distinct company is represented by a node.  A directed edge is placed from shareholder to each company in which they have invested, see \figref{fig:bipartite}.  Note that some companies can be shareholders of other companies, that is, the intersection of sets \{1,2,...,12\} and \{A,B,...,F\} is not necessarily empty. For example in Fig 1, the node 11 would be node A. For example in Fig~\ref{fig:bipartite}, the node $11$ would be node A. Also, we do not have the value of each investment, nor can we be sure that all shareholders are present in our databases.  For those reasons we chose not to try to represent the size of investments e.g.\ through a weight added to the edges.

The boundary of each network is defined by the nationality of the companies; here we study two examples: Turkish companies and Dutch companies. We consider all the shareholders of each company which means that we consider both domestic and overseas shareholders. We will highlight this when analysing the data for companies in the Netherlands.

\begin{figure}[htb]
  \centering
  \includegraphics[scale=0.4]{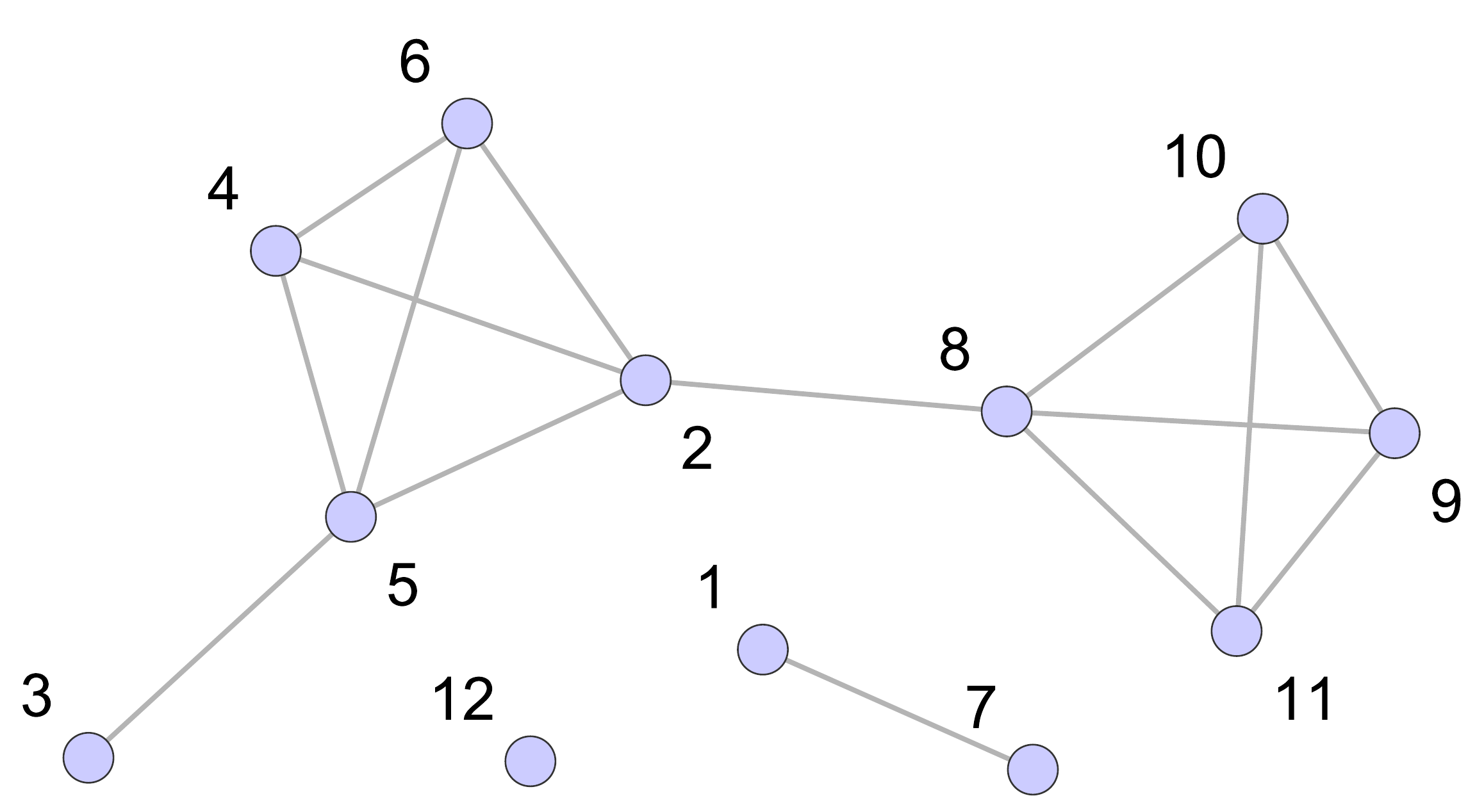}
  \caption{{\bf Shareholder network.} The ``shareholder network'' for the data displayed in \Figref{fig:bipartite}. This is the projection of the shareholder-company network onto just the shareholder nodes. The nodes here are just the shareholder nodes 1 to 12.  An edge indicates that two shareholders have assets in common. For example, shareholders 3 and 5 have both invested in E, therefore, an edge between nodes 3 and 5 exists in this projected graph. Note that a simple network is used, edges have no directions and no weights, and there are no self-edges.}\label{fig:projected0}
\end{figure}

We are particularly interested in the relationships between shareholders implied by their investments.  That is, if two shareholders have invested in the same company they have a common interest and are likely to have similar wider commercial interests. So we will focus most of our work on the analysis of these investor-investor relationships and we do this through a representation of our data in terms of a projection onto just the shareholder nodes. Our ``shareholder network'' has one node for each shareholder, and two different shareholders are connected by an undirected edge if they have both invested in the same company. An example of the shareholder network is shown in \figref{fig:projected0} which is a representation of same data shown in the shareholder-company network of \figref{fig:bipartite}.


An important aspect of our work is that our data also classifies the shareholders to be one of $13$ different types of owner, as listed in \tabref{table:type}. We will use this classification to study how the structure of the shareholder networks depends on the type of owner. It is immediately clear from the numbers of each type that companies in different countries can have very different types of shareholder which already suggests that other aspects of corporate structure will be different.

\begin{table}[htb]
  \centering
  \begin{tabular}{  c | c | p{0.22\textwidth} || r l | r l}
    ID & Shareholder Type          & Description  &\multicolumn{2}{c|}{Turkey} & \multicolumn{2}{c}{Netherlands} \\ \hline \hline
    1  & Venture Cap.\     & Venture capital & 6 & (0.01\%) & 143& (0.07\%) \\ \hline
	2  & Financial Co.\    & Financial company & 133 & (0.23\%) & 7028& (3.6\%) \\ \hline
    3  & Families          & One or more named individuals or families & 53360 & (92.29\%) & 663& (0.34\%) \\ \hline
    4  & Public Co.\       & Publicly listed companies & 1 & (0.002\%) & 1& (0.001\%) \\ \hline
    5  & State             & Public authority State Government & 26 & (0.04\%) & 162& (0.083\%) \\ \hline
	6  & Hedge Funds       & Hedge funds & 1 & (0.002\%) & 16& (0.0082\%) \\ \hline
	7  & Insurance Co.\    & Insurance company & 34 & (0.06\%) & 235& (0.12\%) \\ \hline
	8 & Self Owned        & Self Owned & 1 & (0.002\%) & 1& (0.001\%) \\ \hline
	9 & Private Equity    & Private Equity firms & 18 & (0.03\%) & 282& (0.14\%) \\ \hline
	10 & Corporates       & Industrial company & 4007 & (6.93\%) & 156644& (80.25\%) \\ \hline
	11 & Mutuals           & Mutual, Pension Fund, Nominee, Trust, Trustee & 91 & (0.16\%) & 9563 & (4.90\%) \\ \hline
	12 & Banks             & Banks & 127 & (0.22\%) & 363& (0.19\%) \\ \hline
	13 & Foundations       & Foundation, Research Institute & 15 & (0.02\%) & 20059 & (10.27\%) \\ \hline \hline
         \multicolumn{3}{r||}{TOTALS} & 57820 & (100\%) & 195199 & (100\%)
\end{tabular}
\caption{{\bf Summary statistics for different types of shareholder.} The different types of shareholder recorded in our data as retrieved from the BvD database. The numbers found in our different data sets are in the righthand columns.}\label{table:type}
\end{table}

Finally, in many situations we measure values but we need to see if these are large or small by comparing the results against those in an appropriate null model. Our null model is obtained by swapping pairs of edges in our shareholder graph which maintains the degree of each node \cite{MS02} 
as in the configuration model 
(for example see \cite{MR95}).  However, we only make swaps which maintain the constraint that our edges are always between a shareholder node and a company node as illustrated in \Figref{fig:bipartite}.



\section*{Results}\label{sec:characteristics}

In this section we will look at the results of our analysis of the shareholder network. We will start with some general characteristics of the network before moving on to focus on how the different types of shareholder play different roles in the network as revealed by various measurements.

\subsection*{General Network Analysis}\label{subsec:generalanalysis}

Some key facts for our two data sets and for the shareholder networks derived from them are summarised in \tabref{table:networks_summary}.
\begin{table}[ht!]
\centering
\begin{tabular}{ r || c | c }
Country of Companies & Turkey & Netherlands \\ \hline

No.\ of Companies & 45,831 & 1,157,672 \\ \hline

No.\ of Companies with Information Available & 22,445  & 259,249  \\ \hline

No.\ of Nodes in Shareholder Network & 57,820 & 195,199 \\ \hline

No.\ of Edges in Shareholder Network & 93,439& 133,276 \\ \hline

Degree distribution power-law slope $\gamma$ & 2.6 & 2.7 \\
\hline
No.\ of nodes in LCC & 1,807 & 3,152 \\ \hline
No.\ of edges in LCC & 14,459 &120,935\\ \hline
Clustering Coefficients in LCC &0.87&0.64\\
\hline
Diameter in LCC & 12 & 19 \\
\hline
Average Shortest Path in LCC & 4.50 &4.47\\ 
\end{tabular}
\caption{{\bf Basic information on the two data sets.} Each data set looks at the companies registered in one country and their shareholders from any country for the year 2014. There is no information on many of the Companies as the numbers above indicate. The number of edges in our shareholder network is based on the information available on the shareholding information. The slope $\gamma$ of a power-law degree distribution of similar slope, $P(k) \; \sim \; k ^ {-\gamma}$, is a rough characterisation to illustrate the broad distributions. `LCC' is the largest connected component. }
\label{table:networks_summary}
\end{table}



\begin{figure}[H] 
    \includegraphics[width=\textwidth]{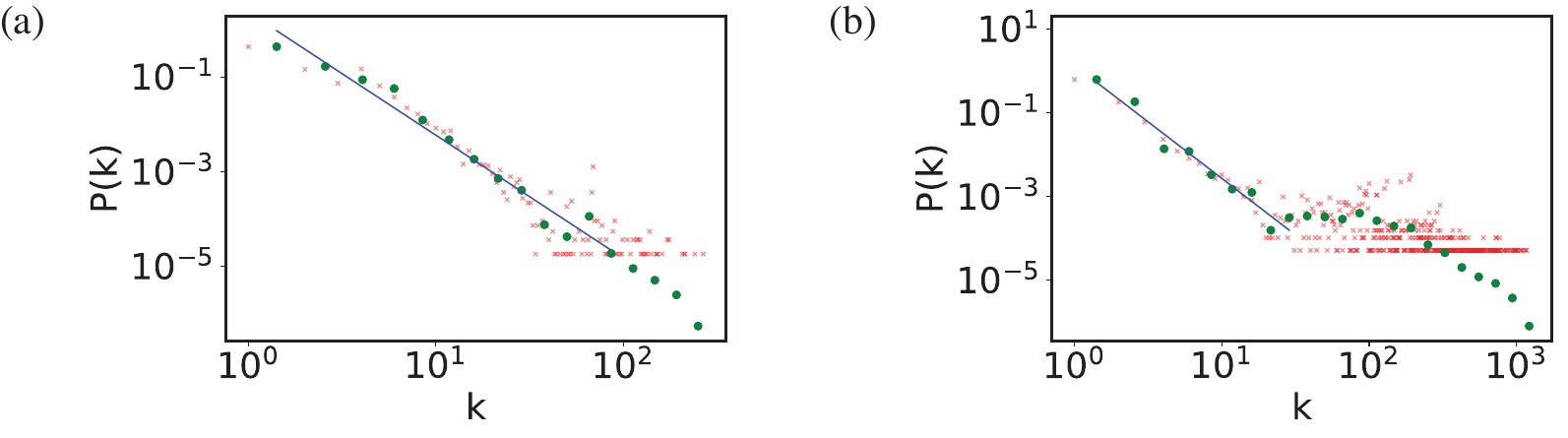}
 \caption{{\bf Plots of degree distributions.} The degree distributions $P(k)$ (the frequency of nodes with degree $k$) against degree $k$ edges on a log-log scale for shareholder networks where the holdings are in (a) Turkish companies, (b) Dutch companies. The red dots are the raw data, the green crosses represent the same data in logarithmic bins, and the blue lines are the best linear fits ($P(k) \; \sim \; k ^ {-\gamma}$) to ranges of $k$ values where we see approximately linear behaviour. The slope of the blue lines, $-\gamma$, is $2.6$ and $2.7$ for Turkey and the Netherlands respectively. A summary of the general statistics of these shareholder networks can be found in \tabref{table:networks_summary}. }
 \label{fig:degree}.
\end{figure}

An important characteristic of any network is the degree distribution, $P(k)$. This may be defined as the probability that a node selected uniformly at random has degree $k$, where the degree of a node refers to the number of edges connected to a node.
The degree distribution for the shareholder networks are shown in \figref{fig:degree}.
The distributions are generally fat-tailed as illustrated by the power-law forms shown in \figref{fig:degree}.

It can be seen from \Figref{fig:degree}, that the distribution of the degrees of the network roughly follows a power-law. The large $k$ tail implies that typically there are a small number of shareholders who have investments in common with large numbers of other shareholders. On the other hand, the small degree part of the distribution indicates that most shareholders have investment in common with only very few others.

The distributions show other interesting features. For the Netherlands there is a distinctive `bump' in shareholders who are related to between twenty and a hundred other shareholders. These appear to be far more common than the trend shown for small degree would suggest. One explanation for this is that some companies have lots of overseas shareholding relationships, for example the French shareholder of the Turkish Tobam Holding Co. With foreign shareholders, it seems likely that most would be large shareholders who are looking to diversify their holdings by looking outside their home company. Including these shareholders has two effects.  First including them increases the total number of nodes in our shareholder network which lowers the distribution $P(k)$ for shareholders investing in companies based in their home country. Secondly, it seems likely that if a foreign shareholder has gone to the trouble of making one investment across borders, it is likely they have made several, so that they are the bump. Put another way this is a boundary effect.  Our large foreign investors will also be linked through foreign firms to small foreign shareholders who only invest in foreign firms, part of the low degree part of the distribution. Those firms are excluded by definition from our data. The fact that the bump is much less pronounced in the Turkish shareholder network suggests that Foreign investment does not play such a big role in this case. As the macroeconomic data suggests that Netherlands has high FDI (International trade and foreign direct investment) indexes both outward and inward, 13.37\% net inflows of GDP. It ranks 9th in the world Netherlands ranks 10th when including Hongkong,China who is number 1 and Turkey ranks 178th. 


\subsection*{Analysis by Shareholder Type}\label{subsec:typeanalysis}

One of the major features of our data set is that we can distinguish between $13$ different types of shareholder as shown \tabref{table:type}. So we will have a look at how various network measures reveal the different roles of different types of shareholder and how that depends on the two countries we are studying.

\subsubsection*{Degree}\label{subsec:degree_cen}

\begin{figure}[H]
\centering

\includegraphics[width=\textwidth]{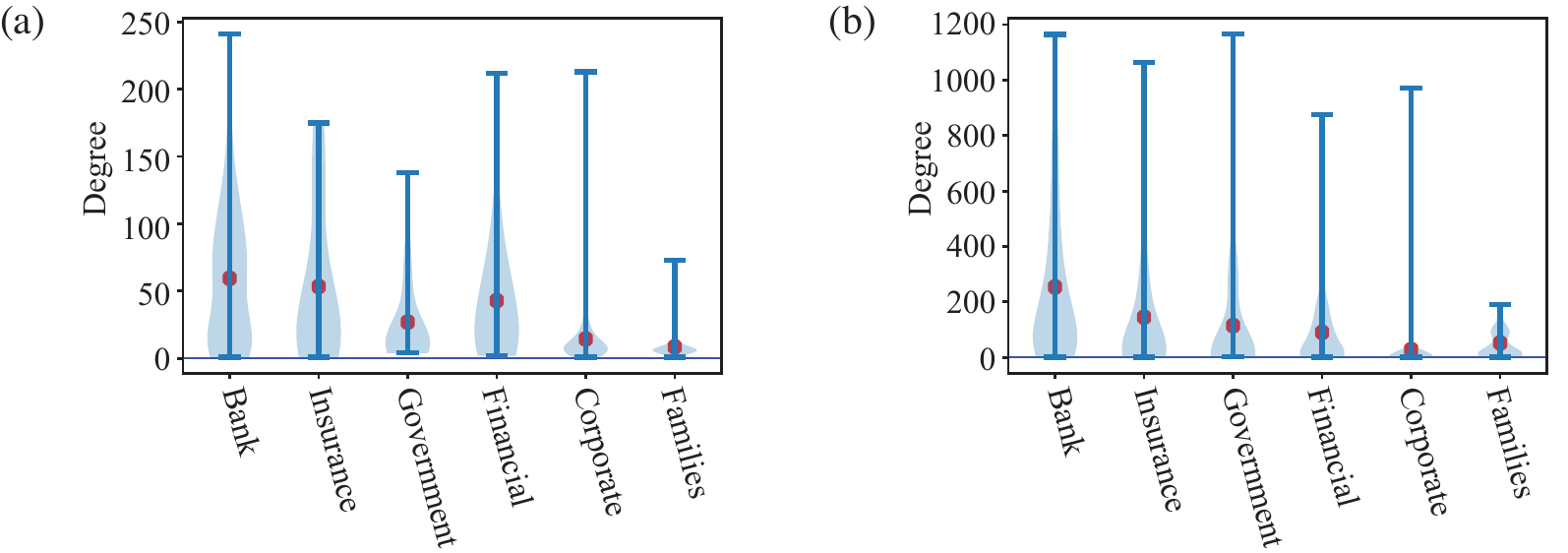}

\caption{{\bf Violin plots of the degree.} Violin plots of the degree of the most common types of shareholders for the largest connected component of shareholder network of (a) Turkish and (b) Dutch companies. There are too few shareholders for other types of investor. This figure breaks down degree distributions into different types of shareholders. We note that in Turkey, the large degrees are contributed by the banks and insurance while in Netherlands, banks' average degree is higher than the other types of shareholders. It means Netherlands' banks co-invested a lot with other shareholders.}
\label{fig:ot_degree}
\end{figure}

The degree of a node is one of most important and simplest centrality measurements.
In our shareholder networks, a high degree of a shareholder may indicate that that shareholder has better contacts within the business community. To evaluate the different role of different types of shareholders in \secref{sec:methods}, the violin plots for the degree of different types of shareholders are shown in \figref{fig:ot_degree}. Violin plots are similar to box plots indicating ranges and additionally show the estimated probability densities of the data at different values and include a marker for the median of the data. 

This degree centrality is straightforward to find from the data, but does full complex network provide more information? We consider several other topological measurements but we start with one of the simplest.  We look at the effect on the largest connected component of removing nodes one at a time, choosing the remaining nodes uniformly at random from those of just one type. The effect of removing different types of shareholder on the largest connected component of the Turkey shareholder network is shown in \Figref{fig:turkey_ccs_remove}.



\subsubsection*{Percolation}\label{subsec:degree_percolation}

In our percolation analysis, we focus on one type of shareholder.  Starting from a given shareholder network, $G(r)$, we choose one node, uniformly at random from the set of shareholders of the given type, and we remove that node and any edges attached to it.  This leaves us with the next network $G(r-1)$ with one less node.  We then repeat the process.  In our case we will start from the the largest connected component of one of our shareholder networks, and then  we will remove the nodes corresponding to one of the more common shareholder types. We will look at the number of components of the sequence of networks $G(r)$ as a function of the number of nodes removed, $r$.

\begin{figure}
\centering

\includegraphics[width=\textwidth]{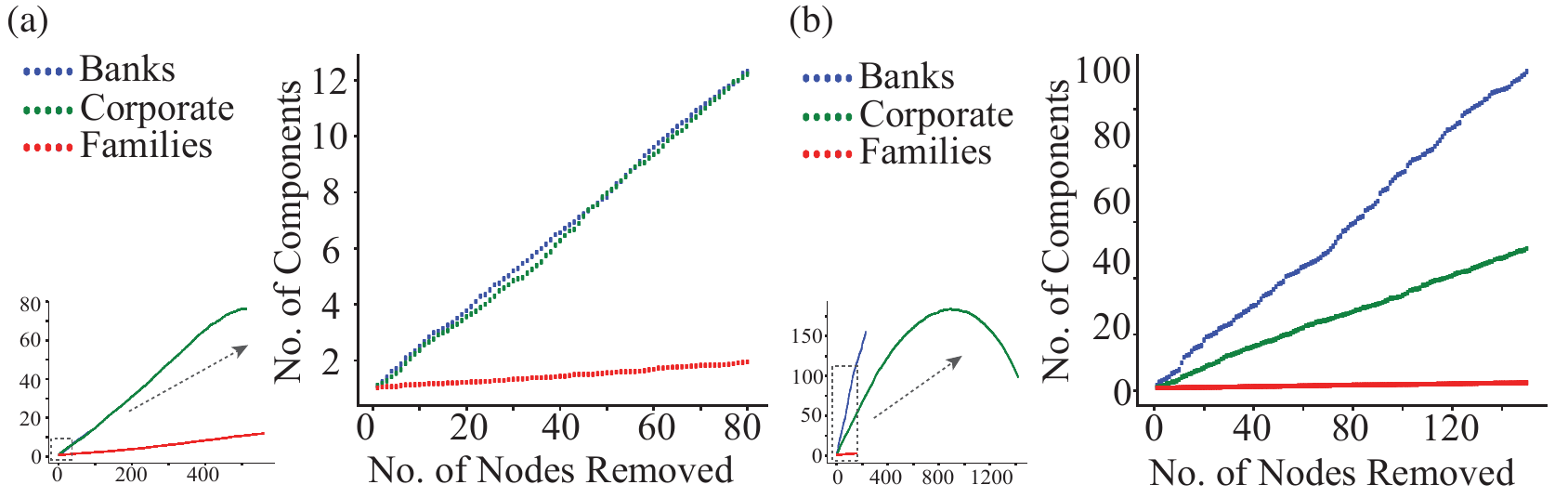}

\caption{{\bf The number of components increases as the number of nodes removed.}  (a) Turkish and (b) Dutch companies. Nodes of one shareholder type are chosen at random and removed one by one from the largest connected component of the shareholder network. Results shown here are averaged over 100 realisations. Blue represents Bank shareholders being removed, green represents Corporate and red represents Families. The larger scale plots display the regions in the dashed boxes of the smaller scale plots to more clearly reveal the behaviour for small numbers of node removals. Note in particular the different role of banks (blue) and corporates (green) in Turkey and Netherlands. The small and big plots share the same axis labels.}\label{fig:turkey_ccs_remove}
\end{figure}

Results averaged over 100 realisations are shown in \figref{fig:turkey_ccs_remove}.  For both countries, we see that the change in the number of components is roughly linear, at least for a relatively large rank of $r$ values, but the slopes are very different. These differences in the percolation analysis give us an insight into the roles of different types of shareholders within the network.

Individual and Family shareholders, which are 92\% of the nodes in Turkish shareholder network while 0.34\% in Dutch shareholder network, seem to have a limited effect on the connectivity of the largest component. This can be understood by the nature of Individual and Family shareholders whom we would expect to have investments in a few closely related companies and so would only be linked to a few closely linked shareholders. That is, it is not surprising if Individual and Family shareholders are poorly connected to other shareholders and are somewhat peripheral to the network.

There are also a large number of nodes corresponding to Corporate  shareholders, just under 7\% in Turkish shareholder network and 80\% in Dutch shareholder network, and their average degree is again not high. Being different from Family shareholders, removing this type of shareholder breaks up the giant component much more quickly. So Corporate shareholders seem to be important in bringing together smaller components. For instance, in the real world, companies involved in mergers and acquisitions are likely to bring together different bodies of interest.

Since Banks shareholders often invest in a large number of different assets, in terms of the shareholder network they are going to be responsible for providing a path in the shareholder network between many different types of shareholder. This central role is reflected in the fast rate at which their removal breaks up the largest connected component.

\subsubsection*{Investor Assortativity}\label{subsec:assort}

Different types of investors mix with other types of investor to different extents.  This can be measured by looking at the the assortativity in the types of investor at the end of each edge. The covariance of the investor labels associated with the two ends of each edge is defined as\tnote{For example see section 7.13.1 p.222 in \cite{newman2010networks}.}
\begin{equation}
    \mathrm{cov}(\tau,\sigma) = \frac{1}{2m}\sum_{ij} \left(A_{ij} - \frac{k_{i}k_{j}}{2m}\right)T^\tau_{i}T^\sigma_{j} \, , 
\end{equation}
where $T^\tau_{i}$ is one if vertex $i$ is an investor of type $\tau$ and this is zero otherwise. Here $m$ is the number of edges. To measure assortativity, we look at the diagonal elements, $\mathrm{cov}(\tau,\tau)$, to see if edges have the same type of investor at both ends.  If we sum these and normalise by the variance in the labels we arrive at the investor type assortativity coefficient $r$
\begin{equation}
    r = 
    \frac{\sum_{ij}\left(A_{ij}-k_{i}k_{j}/2m\right)\delta(\tau_{i}, \tau_{j}) } {\sum_{ij}\left(k_{i}\delta_{ij}-k_{i}k_{j}/2m\right) \delta(\tau_{i}, \tau_{j}) } \, ,
\end{equation}
where\tnote{So $\delta(\tau_{i}, \tau_{j} )  = \sum_\tau T^\tau_{i} T^\tau_{j} $.} vertex $i$ is an investor of type $\tau_i$. 
This varies in value between a maximum of $r=1$ for a perfectly assortative network, in which each investor is only linked to investors of the same type, while the minimum value of $r=-1$ indicates a perfectly disassortative network. A value of $r=0$ implies that the types of investor at the ends of edges are uncorrelated.

\begin{figure}
\centering
\includegraphics[width=\textwidth]{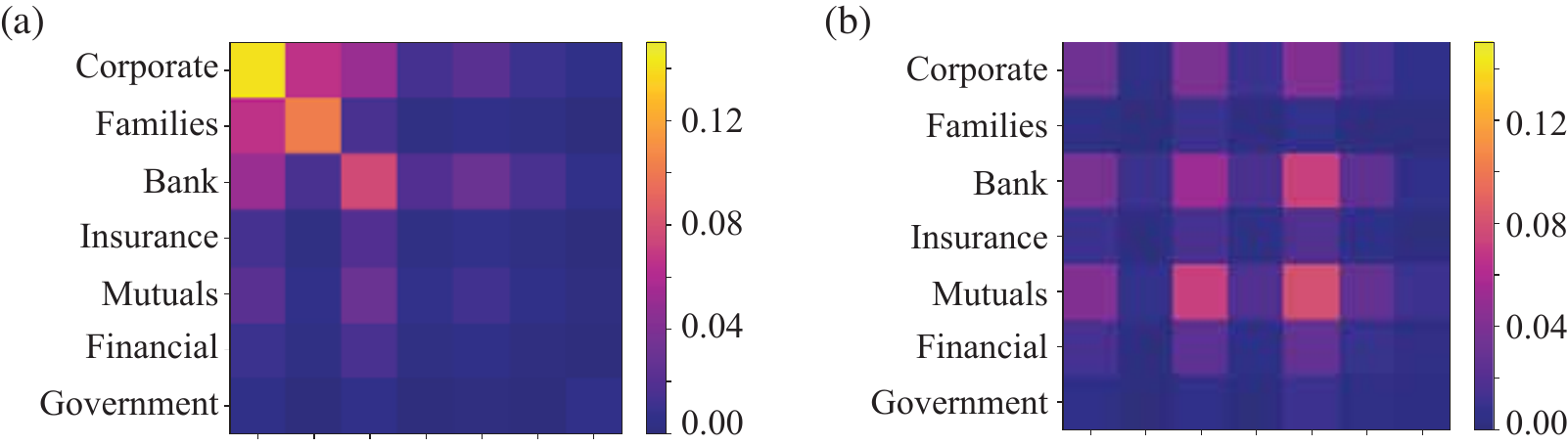}
    \caption{{\bf Heatmaps of degree assortativity.} The degree assortativity in the LCC for different investor types. It has been normalised for the counts of pair of each edge in LCC. The assortativity coefficient $r$ is $0.17$ for Turkey and $0.0081$ for Netherlands. The colour bar is set at the same scale}
    \label{fig:assortativity}
\end{figure}

Our results are shown in \figref{fig:assortativity}. In the case of Turkey, the Corporate shareholders have a high assortativity, indicating that they prefer to invest with each other. Likewise Family investors in Turkey prefer to invest with each other.  Despite this similarity in terms of assortativity, other network measures will show also Family and Corporate investors in Turkey play very different roles.
For the Netherlands, the only standout feature is that Mutuals and Banks prefer to invest with each other, suggesting that in the Netherlands these investors are tightly linked.


\subsubsection*{Diversity of Neighbours}\label{subsec:diversity}

\begin{figure}[H]
\centering

\includegraphics[width=\textwidth]{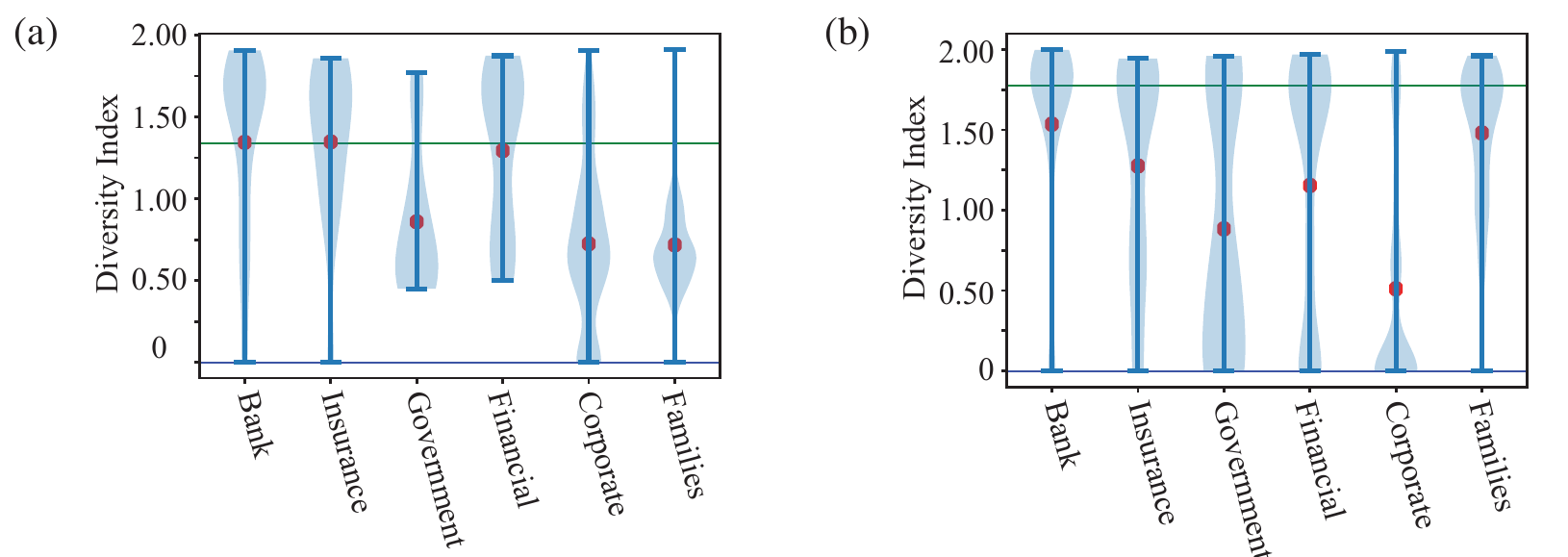}

\caption{ {\bf Violin plot of diversity index of selected types of shareholders.}  (a) Turkey and (b) Netherlands. The information of other type is not listed here because due to the limited information available and the limited amount of the data. The blue space is the diversity index density estimation and compared with a null index (indicated by a green line) which is define as $d^\mathrm{null}$ in Eq. \eqref{dnulldef}.}\label{fig:ot_div}
\label{fig:diversity}
\end{figure}

Interestingly, while Banks and Corporate shareholder nodes are important in maintaining the connectivity of the shareholder network, there is an important difference in their share holding patterns. To see this we turn to a measure of the diversity of the neighbours, $d(i)$, of a node $i$ in terms of the  different types of shareholders. Our measure of diversity of a node $i$ is defined as:
\begin{equation}
 d(i) = -\sum_{\tau}\frac{k_{i}(\tau)}{k_{i}}\ln\left(\frac{k_{i}(\tau)}{k_{i}}\right),
 \quad \mbox{ where } \quad
 k_{i} = \sum_{\tau}k_{i}(\tau)
 \, .
\end{equation}
Here $k_{i}$ is the degree of node $i$ and $k_{i}(\tau)$ is the number of neighbours of node $i$ which are of type $\tau$. If the neighbours of a node $i$ are all of the same type, say $k_i(\tau)=\delta_{\tau,\tau_0}k_i$, then $d(i) = 0$. However, if the neighbours of node $i$ are all of a different types, $k_i(\tau)=1$, then diversity would be $\ln(k_i)$. To make a suitable comparison, we find the expected measure of diversity $d_\mathrm{null}$ given the distribution of labels in each data set, that is
\begin{equation}
    d_\mathrm{null}
    =
    - \sum_{\tau} \frac{N(\tau)}{N} \ln\left(\frac{N(\tau)}{N}\right)
    \label{dnulldef}
\end{equation}
where $N(\tau)$ is the total number of nodes of type $\tau$ and we have $N = \sum_{\tau}N(\tau)$.
The null model diversity measurements indicates the global diversification of different types of shareholders within one country. In Fig~\ref{fig:diversity}, we see that in terms of the classification scheme used in our data, the Netherlands has a much more diverse set of shareholders than Turkey.
If a node's diversity is lower than this expected diversity value, this indicates attraction of certain types of shareholder to the same investments. On the other hand, if a node's diversity is higher than might be expected at random, this indicates that some types of shareholders repel each other, as the probability of them co-investing is lower than expected.

Diversity indices for Turkey correlate roughly with degree for different types of shareholders. The exception is the Family shareholder type whose diversity index is the lowest and below the global average model, indicating these shareholders tend to invest with a very limited range of co-investor types. One explanation is that many Family type shareholders only invest  with the same type of shareholders and perhaps, in many cases, these connections reflect real social and family ties. We will see further evidence for this view in other measures. The Netherlands' diversity index is interesting as most of the shareholder types have mean diversity measures below the global diversity measure, showing some tendency for Dutch shareholders to invest with a limited set shareholder types. Overall though the values of diversity measurement of the same type in two countries are similar, implying that in terms of diversity, the behaviour of different types of shareholder is similar in different countries, except the type Families.


\subsubsection*{Betweenness Centrality}\label{subsec:bet}

Another way to study how the roles of different types of shareholders vary in our network, it is useful to look at how \textit{betweenness centrality} values vary.
Betweenness centrality measures the number of shortest paths passing through a node. In the context of our shareholder network, the shortest path can be interpreted as the the minimum number of common assets that connect other two shareholders as each edge represents a shared asset between a pair of shareholders. The interpretation is that the higher the betweenness the more likely they will be central to the process of connecting other shareholders making them more important to other shareholders.

\begin{figure}[H]
\centering

\includegraphics[width=\textwidth]{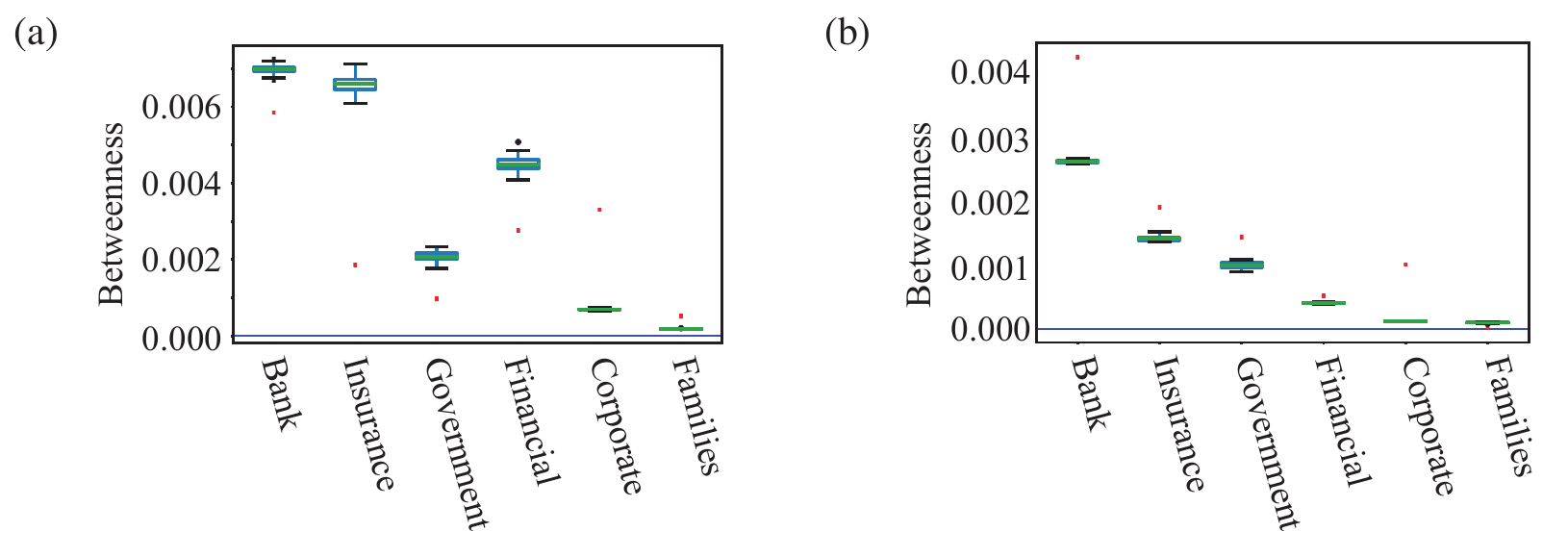}

\caption{{\bf Plots of Betweenness.} The mean betweenness values for different types of shareholders in the largest connected component of shareholder networks, (a) for  Turkey and (b) for the Netherlands. The red dots are the real data and the box plots for the results obtained from 100 degree preserving null models. We note that most betweenness values for Turkey and Netherlands are significantly different from the randomised networks, some types are lower and some types are higher. That means that there is significant network structure on larger scales and the properties are not just controlled by the degree.  }\label{fig:ot_bet}
\end{figure}

In order to see if these betweenness values are significantly high or low, and so to see if this measure gives different information from the degree, we compare our values against those in our null model in which the edges are swapped but the degree of each node is unchanged.
We create 100 different null models and use these values to create the boxes in \Figref{fig:ot_bet} alongside the results obtained from our data.

In \figref{fig:ot_bet}(a) we show the betweenness centrality values for each type of shareholder in the Turkish shareholder network, as well as the results from our null model. In this case, Banks always have the highest average betweenness and highest maximum betweenness. This implies that a high percentage of shortest paths go through Banks which in turn means that Banks can play a pivotal role in linking other shareholders. As these are key instruments for providing investments in firms, this is not surprising. However, this network measure confirms our intuition
and hence we see these companies are fulfilling their role in the economy.

However, nodes representing Banks, Financial and Insurance companies are much less central than you expect from the null model.  These companies have a  high degree yet their betweenness is lower than one might expect.  So Banks,  Insurance and Financial companies are still very central in the network but they are much less effective in brokering connections than we would expect from their degree value, the message from their betweenness values in the null model. What this suggests is that in Turkey, Banks, Insurance and Financial companies are investing in a narrower range of companies than they could.



The State organisations in Turkey are also less central than expected, suggesting their involvement is constrained by some issues, e.g.\ political or legal constraints limiting involvement to certain key sectors or to just a few larger firms.

On the other hand while the largest number of shareholders in Turkish companies are the Family type shareholders, this type of shareholder has the lowest average betweenness. This is consistent with what we found from both the low degree of most Family shareholders but also from the diversity measures that the focus of many Family type shareholders may be framed within a social and family setting.  Another explanation may be that the size of these investments may be smaller, again biasing their involvement to smaller firms. The picture is that the investments made by Family type shareholders are peripheral to the large scale shareholding structure in Turkey. Our results on community detection in section~\ref{subsec:comm} will support this view.


When we compare the results for Turkey against those for the Netherlands, we see two big differences as the Banks and Insurance companies investing in Dutch companies are much more central than found in the null model, the opposite of our result for Turkey.

\subsubsection*{Closeness Centrality}\label{subset:close}

While betweenness centrality indicates how a node may control the important communication pathways between shareholders, closeness centrality indicates how easy it is for each shareholder to reach any other shareholder. The closeness of a node is the inverse of the sum of the shortest path distances from that node to all other nodes.  The larger the closeness of a node, the shorter the distances to other nodes and so in general there are fewer message transmissions, less information is lost, communications will be faster and generally will cost less.

In the context of our shareholder network, information can be related to opportunities to buy new assets or to sell existing ones. Since the network is highly interconnected, a failure in one sector can have repercussions in another so the earlier a shareholder hears about potential problems, the more successful they are likely to be.

\begin{figure}
\centering

\includegraphics[width=\textwidth]{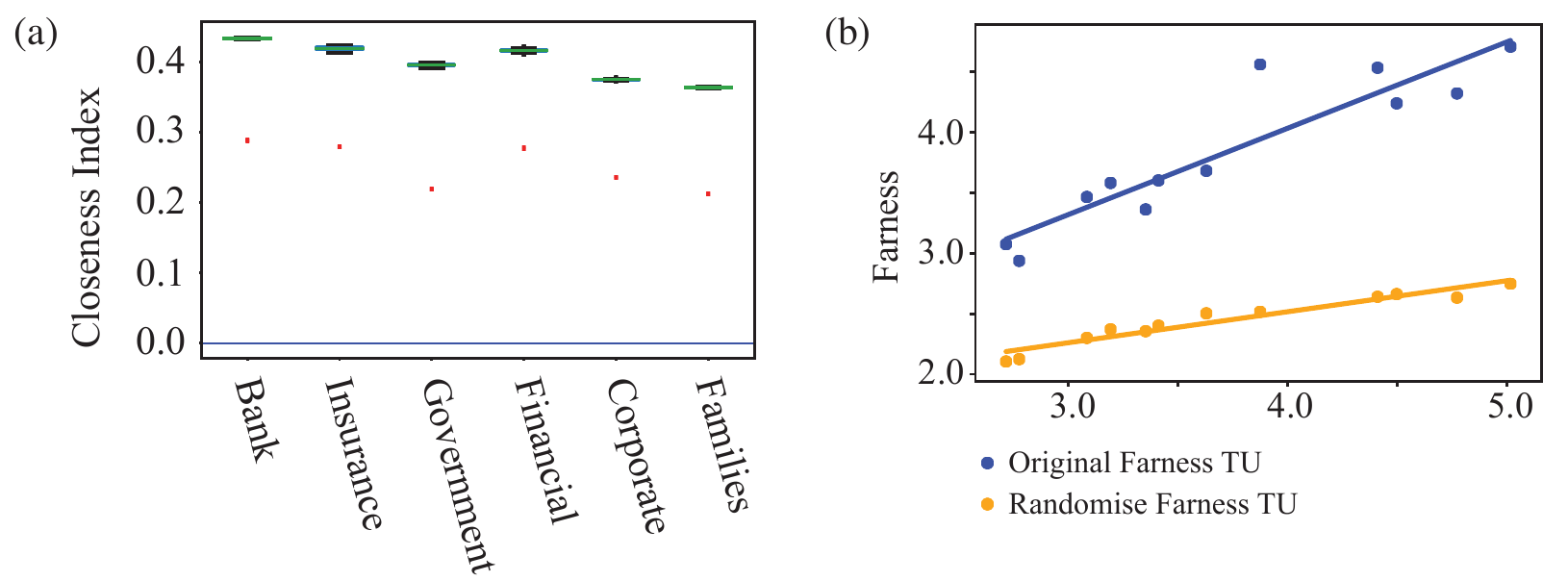}
 
\caption{{\bf The average closeness indices for different types of shareholders in the LCC of Turkish shareholder network.} In Figure (a) we show the results for each shareholder with the red dots for the original data while the box plots are for the randomised data.  In (b) each point shows the  average `Farness' (the inverse of closeness) of one shareholder type against $\log(N/k)$, where $k$ is the average degree of nodes of that type. The higher blue points are for the original data, the lower orange points are for the randomised network.  
The lines in (b) are for a linear fit to the points. The slope of this fit to the original data is $0.71$, $0.26$ for the randomised network and the theoretical value in a random branching model is $0.24$.}\label{fig:turkey_close}
\end{figure}

\begin{figure}[H]

\includegraphics[width=\textwidth]{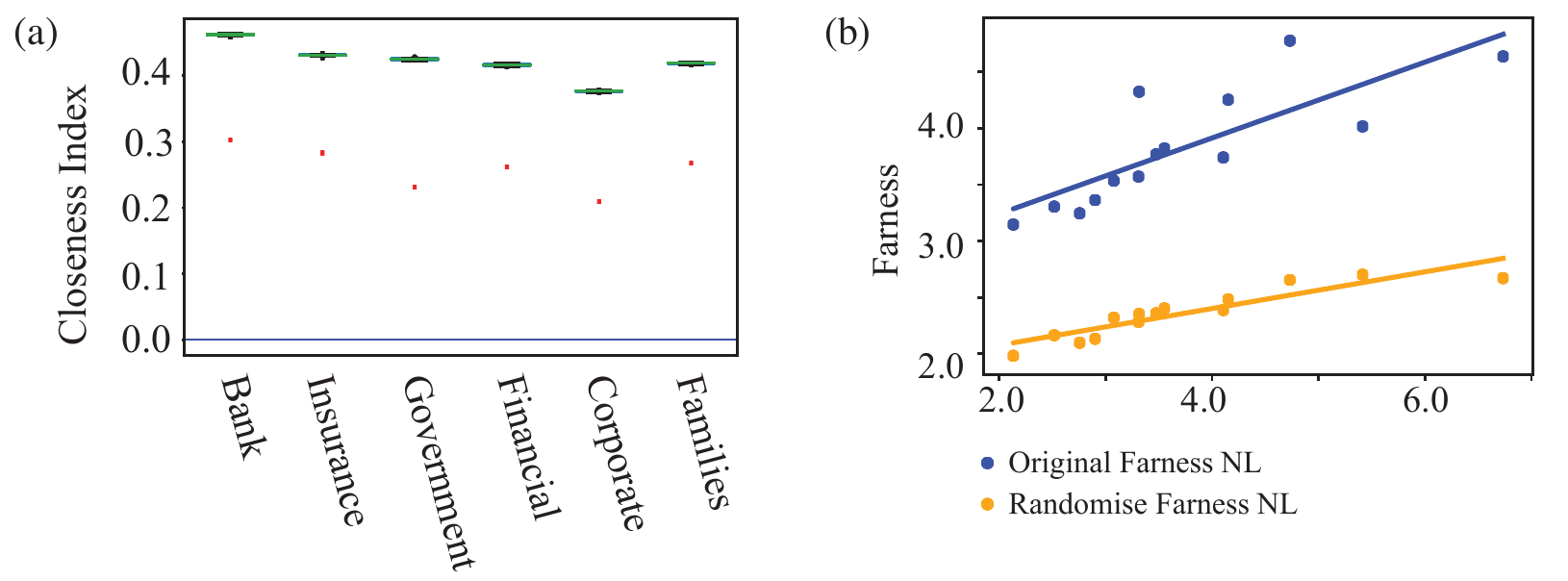}

\caption{
{\bf The average closeness indices for different types of shareholders in the LCC of Dutch shareholder network.} In Figure (a) we show the results for each shareholder with the red dots for the original data while the box plots are for the randomised data.  In (b) each point shows the  average `Farness' (the inverse of closeness) of one shareholder type against $\log(N/k)$, where $k$ is the average degree of nodes of that type. The higher blue points are for the original data, the lower orange points are for the randomised network.  
The lines in (b) are for a linear fit to the points. The slope of this fit to the original data is $0.34$, $0.16$ for the randomised network and the theoretical value in a random branching model is $0.17$.
}
\label{fig:netherlands_close}
\end{figure}

To see how the closeness varies for the different types of shareholders, we use the edge swapping techniques of our null model to make comparisons. In \Figref{fig:turkey_close} and \Figref{fig:netherlands_close}, we see that for the both shareholder network that the average closeness of each shareholder type is lower than in the randomised model and that this shift is similar for all types of shareholder.  This tells us two things.

First that there is additional structure in the real world over the null model.  That is revealed by the change in closeness between real data and our null model. The fact that null model always has higher closeness can be explained if there are many peripheral nodes, many nodes which are at a large distance from most other nodes.   Our null model will bring in lots of `short cuts' to/from these peripheral nodes, the distances to these peripheral nodes drop and so the null model closeness values are higher. 
Put another way, connections in the real shareholder network mean that communication within the network is not as efficient as it could be. 

For Turkey, most nodes represent Family type nodes and these have the lowest closeness, suggesting they form the bulk of the peripheral nodes. For the same reason, it is the corporate shareholders who are peripheral for The Netherlands.  Conversely, in both cases, though there are few banks, these have high closeness indices suggesting they are not part of the periphery.

Overall, the average closeness values for all shareholder types appear to be dominated a large set of peripheral nodes who are less well connected to the global network than they could be, as the randomised networks show.
The only significant difference in closeness values between the different shareholder types is a reflection of their average degree, making this centrality measure somewhat redundant when looking at averages across groups of node. That is not to say that closeness is not useful.  For individual shareholders, a comparison with the typical behaviour, in terms of degree and closeness, can lead us to find interesting outliers (say low degree, high closeness) worthy of investigation for a given context.

\subsubsection*{Community Detection}\label{subsec:comm}

\begin{figure}[H]
  \centering

  \includegraphics[width=0.55\textwidth]{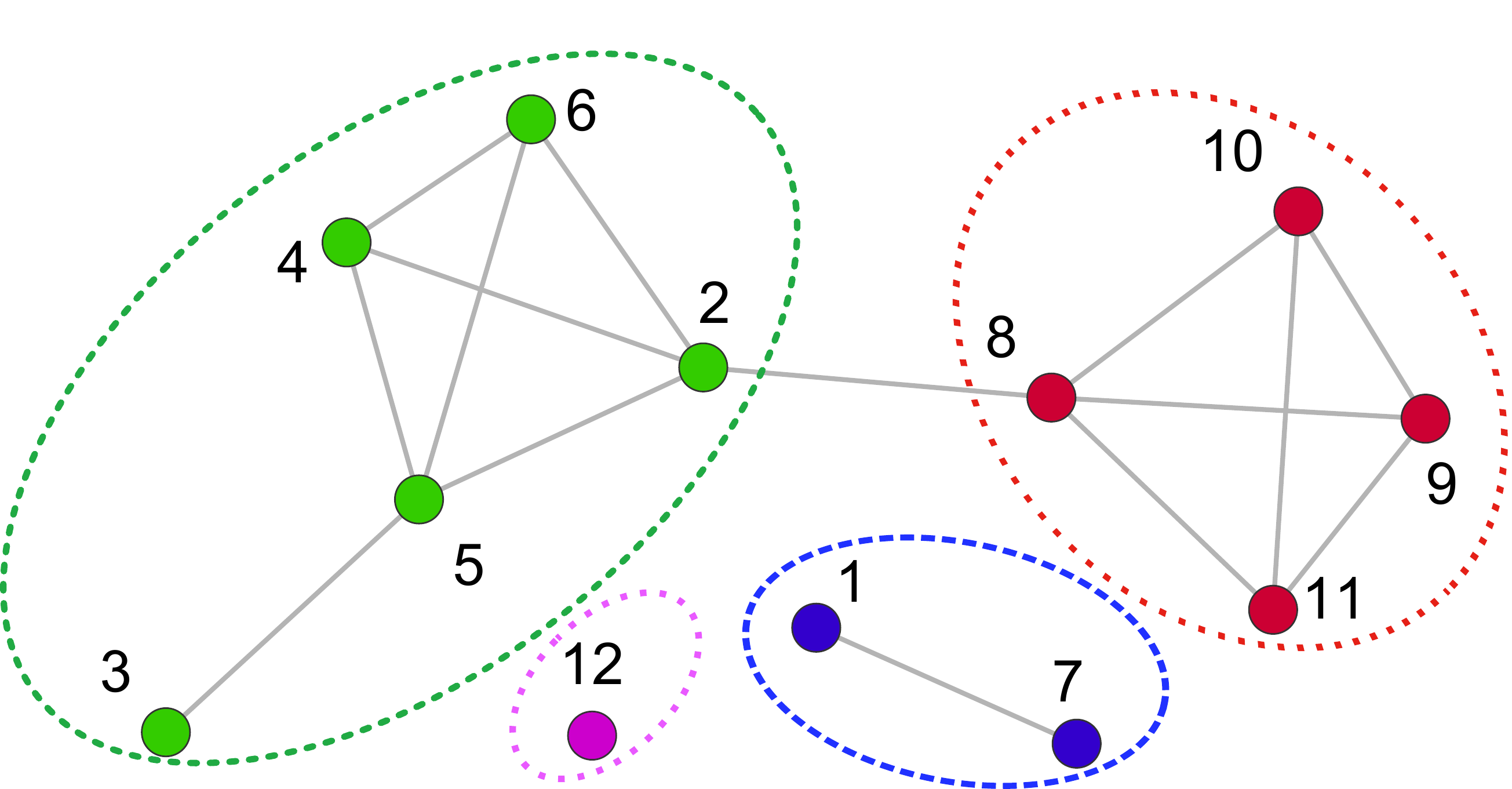}
  \caption{{\bf Illustration of communities in a network.} The same projected graph as in \Figref{fig:projected0} from the network graph shown in \Figref{fig:bipartite}. The different colours label them as different communities which are the structural characteristics in the network science context. As in the graph, 1-12 shareholders are categorised into 4 communities, 12 belongs to one community, 2,3,4,5,6 belong to another community, 8,9,10,11 are in the third community and 1,7 are in the fourth community.
   }
  \label{fig:projected_colored}
\end{figure}

The shareholder network can also show us if the common shareholdings reveal large scale `communities' within the shareholders, more than the labels in the data record.
Such communities in the projected network show us groups with common interests. An illustration of communities is shown in Fig~\ref{fig:projected_colored}. Looking at these groups can tell us something about the diversity of shareholders for each corporate or the centrality of shareholders of the whole economy, which has been discussed in \secref{subsec:owner}.

To do this we use community detection methods to look for groups of nodes which typically have more connections between themselves than one might expect, and/or fewer connections to nodes outside a community. Two popular algorithms have been used here to detect the communities: the Louvian method \cite{BGLL08} and Infomap \cite{rosvall2008maps}, see S1 Appendix for more details on these methods. We construct a distribution for the size of the communities we find. Using two approaches gives us a handle on the uncertainties in this process and we look to see to which community each node belongs to for the two methodologies separately. Some statistics are listed in \tabref{tab:community}.


\begin{table}[H]
\centering
\begin{tabular}{l||c|c||c|c}
Country & \multicolumn{2}{c||}{Turkey} & \multicolumn{2}{c}{Netherlands}\\ \hline \hline

Methodology            & L       & I      & L       & I              \\ \hline
No. of communities     & 21,175  & 21,270 & 182,263 & 182,375    \\ \hline
Avg.~community size    & 2.73    & 2.71   & 1.07    & 1.07   \\ \hline
Max.~community size    & 1169    & 190   & 1532    & 1384    \\ \hline
Avg.~CS excluding single nodes    & 3.03    & 3.01   & 3.03    & 2.98   \\ \hline
\end{tabular}
\caption{{\bf Statistics of the communities.} Communities are found in the shareholder networks derived from the two data sets using the two different methodologies, Louvain (L) and Infomap (I). `Avg.\ community size (CS)' is the average number of shareholders in one community. The average community size is defined by as the number of shareholders divided by the number of communities, where the number of communities include the communities whose community size is 1. The average community sizes excluding single nodes are: 3.03 and 3.01 for Turkey and 3.03 and 2.98 for Netherlands} \label{tab:community}
\end{table}

\begin{figure}[H]
\centering
\includegraphics{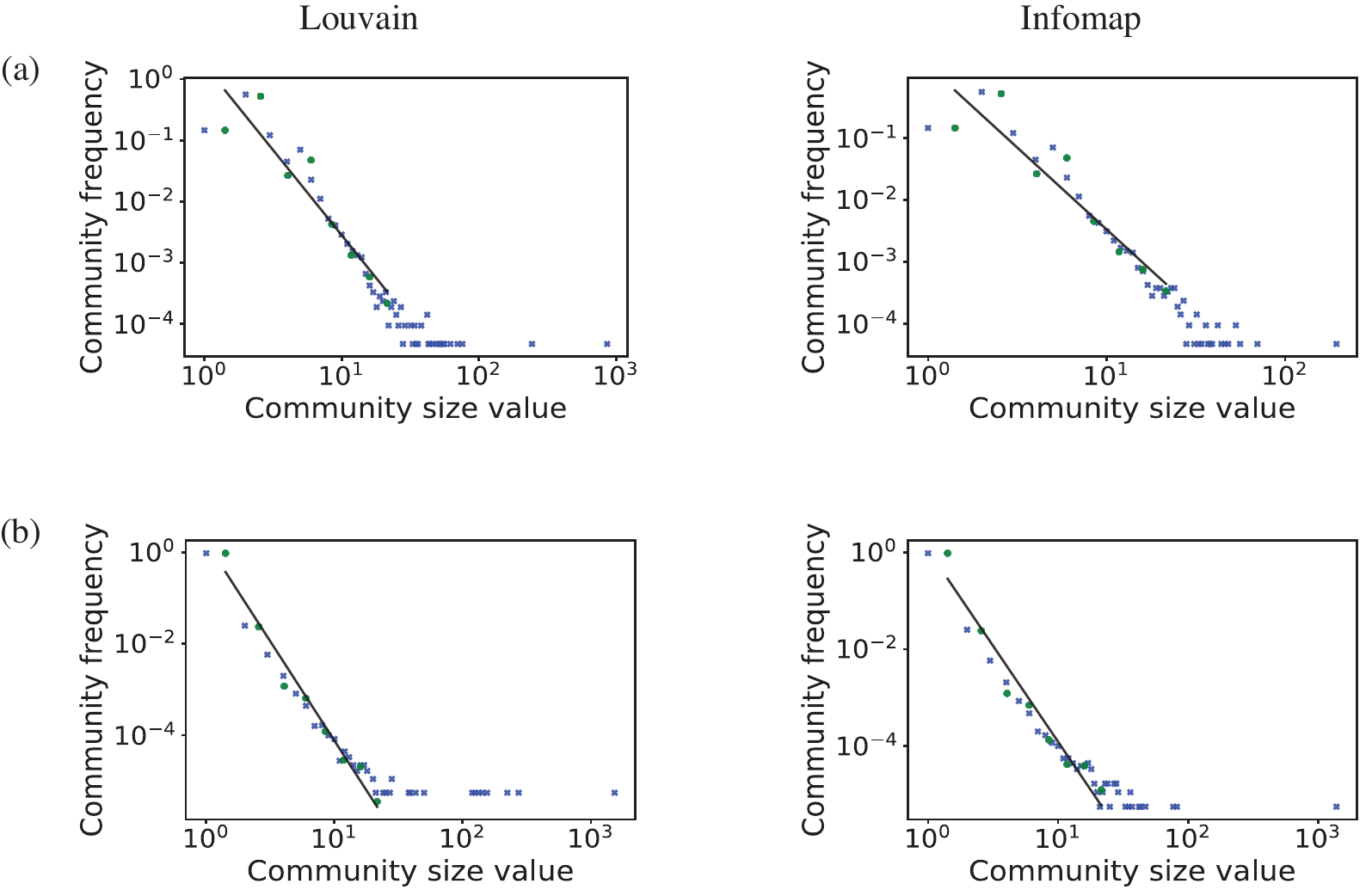}

\caption{{\bf Community size frequency distribution.} Community size frequency distribution for (from top to bottom): (a) Turkey and (b) the Netherlands. The figures are plotted on log-log scale. For each country we show the results from two methods; Louvain on the left and Infomap on the right. The blue cross represents the data, the green dot represents the data binned using a logarithmic binning, and the black line is a linear fit to the binned data.}
\label{fig:community_hist}
\end{figure}

In \Figref{fig:community_hist}, we show the community size distribution on a log-log plot for Louvain and Infomap method for each country

The community size distributions are clearly fat-tailed and power-laws, indicated by the straight lines on the plots, capture most of the behaviour. These distributions show that the vast majority of communities are small, typically three or four shareholders. These are simply disconnected components of the shareholder graph created when a small number of shareholders invest in the same one or two companies. Their shared connections mean these shareholders form a strong community.

The tail of these community size distributions in \figref{fig:community_hist} shows that there are a small number of large communities representing shareholders have cross-invested in each other's investment portfolio but in a way that is highly correlated. Comparing against our null model, we find that such correlated cross investment, the fat-tail of the community size distribution, disappears after our edge swapping. This again shows that the shareholder network is not like a random graph, it has significant structure which reflects a non-trivial way in which these connections are made.

\begin{figure}[H]

\includegraphics[width=1\textwidth]{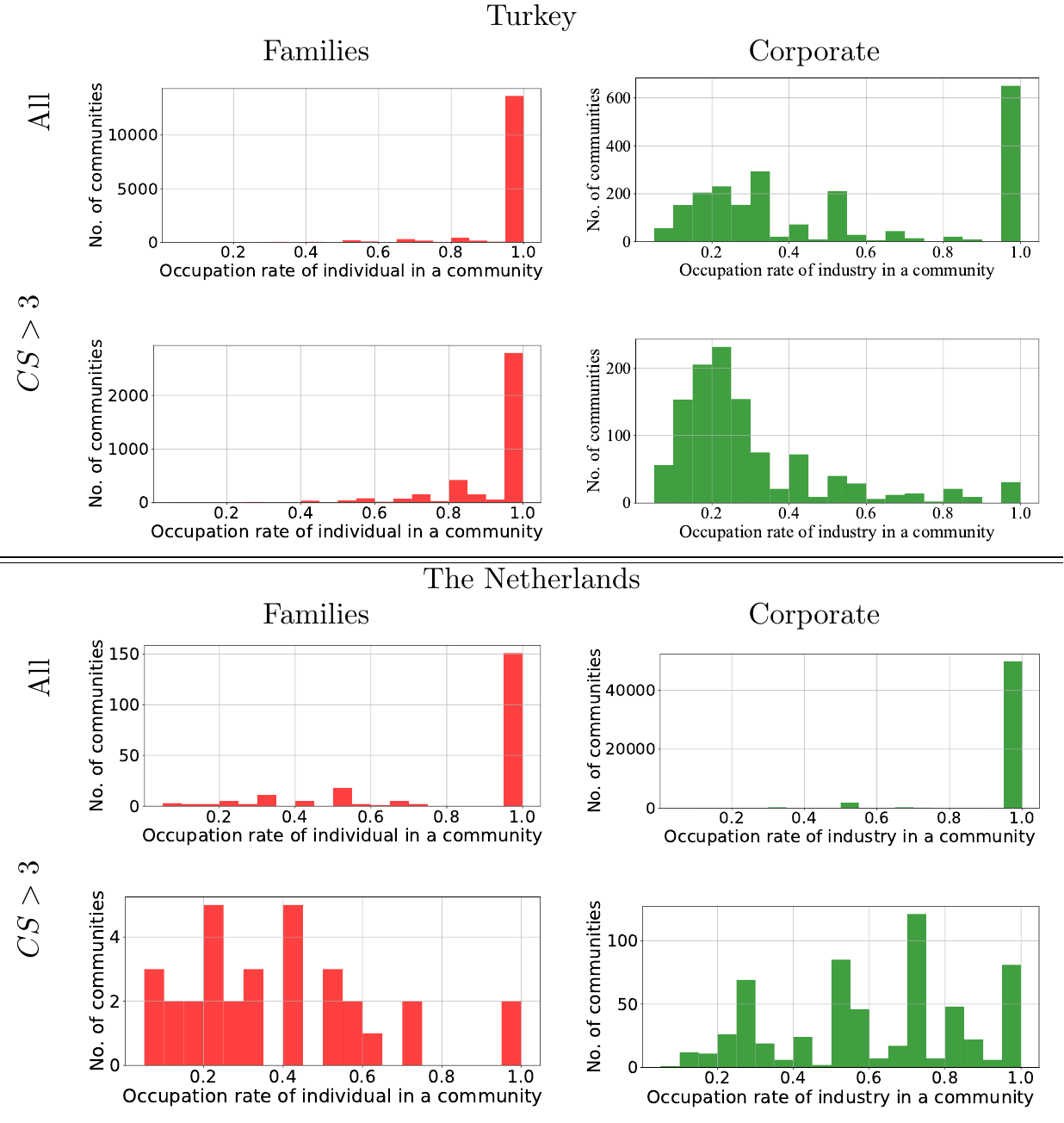}

  \caption{{\bf The number of communities found with a given fraction of one type of shareholder.}  The communities are found with Infomap in the shareholder network for Turkey and Netherlands. On the left we have the fraction of Family shareholders in different communities while on the right we have the fraction of Corporate shareholders in each community. The figures in first row includes community of all sizes.  The fat-tailed distribution means this is dominated by the large number of small communities, and these are almost always of a single type of shareholder, hence the peak at 1.0. The second row shows the same analysis done when we exclude small communities which have three or less nodes (CS = community size).  Similar analysis for the Louvain community detection method is given in the S1 Appendix.}
  \label{fig:2d_sec}
\end{figure}

To see what we can learn from these community structures we look at the kind of shareholder we find in each community using the classification of our $13$ type of shareholders shown in \tabref{table:type}.
We will take Turkey as an example. In this case two types of shareholder dominate: the Industrial investor (Industrial companies)  and the Family investor (`one or more named individuals or families'). In any one community, we look at the fraction of shareholders of these two common types in the different communities.
The distribution for each of the two common types of shareholder are shown \Figref{fig:2d_sec}.


It can be seen from \Figref{fig:2d_sec} that Individual or Family shareholders behave quite differently from the Corporate shareholders in these two countries. Individual and Family shareholders prefer overwhelmingly to invest in companies with the same type of shareholder.  One explanation is that this preference for other Family type owners reflects genuine family ties in the social sense. In general though, individual or family shareholders tend to bond together and exclude other types of shareholders. By way of contrast, we can see that Corporate shareholders are much more happy to share control with other types of shareholders.

However, if we exclude the large number of small communities, those of size one, two and three shareholders, the number of communities we do not see much change in the pattern of ownership for those with family type shareholders in Turkey, there is not much change in the types of shareholder in these larger communities, Family type shareholders prefer to share control with other Family type shareholders.  On the other hand, we do see that larger communities containing industrial companies are far more likely to have mixed types of investor. This phenomenon can be explained by the fact that individual or family shareholders are mainly in small isolated communities which is created by few common investments. In contrast, the corporate shareholders in Turkey and Netherlands appear in both large and small communities. In small communities, they do not invest with other types of shareholders, while in large communities, they have relatively low occupation rate. Further detailed comparison of largest connected component has been provided in S1 Appendix.

The co-invested structures of individual or family are small simple and pure and this supports a picture of the controlling power of family social unit as discussed in the work of Villalonga and Amit \cite{villalonga2010family} and that of Yurtoglu~\cite{yurtouglu2003corporate}.

\section*{Conclusion}\label{sec:discussion}

A core strength of network science is its ability to model relationships between individuals while allowing us to capture the structure of the network on bigger scales and to find the impact of this structure on the individuals within the group. Here we have used this approach to build a network of shareholders and their relationships, as defined by common share holdings. The key to this is to be able to construct the network from real data sources which are difficult and expensive to obtain and require extensive cleaning.  We have shown this can be done by producing networks for two different countries. An important aspect of our networks is that we retain information on the types of shareholder involved so that we provide a new perspective on the roles of these different types of shareholder.

One network feature of note is the way that closeness centrality is found to be of little use as it is highly correlated with degree, in fact linear correlation between the inverse of the closeness of a node with the logarithm of the degree of that node. This stems from the fact that much of its behaviour is dominated by the network at large distances from any node where the network, at least statistically and in terms of the shortest path routes, acts like a random graph and so like a random tree.

Our network analysis has highlighted several features in the data.  One particular one is that the role of the individual or family investor in Turkey is far more peripheral that found in the data for the Netherlands.  We have seen this in percolation, diversity, and betweenness measurements and in the makeup of the network communities. Likewise, the properties we have found for the Corporate shareholder in Turkey suggest they form a central core for that network. This observation suggests that the  core-periphery paradigm \cite{SK79} could be useful here perhaps using one of many ways to quantify the concept such as \cite{BE00,DDP13}. We leave this for later work.
Another observation has been the way what are termed Bank shareholders seem to have a different role in the two countries, more important than other types shareholders in the Netherlands and Turkey.

Looking ahead, one application of our methods in the context of finance would be to evaluate the risk in such networks.
Our percolation measurement illustrates the principle.  By removing nodes at random we see how the network has different vulnerabilities to random failures in different types of shareholder. We could also use our network to see how the loss of confidence of one shareholder might spread through the network, effecting the price of different companies in different ways.  This would illustrate how negative (or positive) effects travel through the networks which will give rise to the systematic risk.

Another future direction is to look at similar data sets from different time periods and to see how the network changes over time. Can we find a model of the behaviour of shareholders at the microscopic scale which shows the macroscopic evolution of the network such as the phenomenon of takeovers?


\clearpage



\section*{Acknowledgement}
The authors would like to thank Eduardo Viegas, Henrik Jensen, Tarun Ramadorai, Yangshen Yang and Nanxin Wei for useful comments.



\clearpage
\renewcommand{\thesection}{\Alph{section}}
\setcounter{section}{0}
\renewcommand{\thesubsection}{\Alph{subsection}}

\section*{Appendix}\label{sec:appendix}

\subsection{Network Definition}

We will set up the basic notation and definitions of the networks used in this work.  We have a set of \paperdefn{shareholders} $\Scal$, labelled $s_{i}$, who hold shares in one or more \paperdefn{companies}, the set $\Ccal$ labelled $c_{j}$ etc. In addition, each shareholder $s \in \Scal$ carries a \paperdefn{type} label $\tau(s) \in \Tcal$ where $\Tcal$ is the set of fifteen different labels as given in the main paper. 

It is sometimes convenient to indicate the subset of shareholders of one particular type so we use $\Scal_\alpha$ to indicate the set of shareholders of type $\tau \in \Tcal$
\begin{equation}
  \Scal_\alpha = \{ s | s \in \Scal, \tau(s) = \alpha \}.
  \label{Scalalphdef}
\end{equation}

We can use our data to define a \paperdefn{Corporation-Shareholder network}, $\Bcal$ in which the set of nodes, $\Vcal_B$, are the union of the set of shareholders and companies, $\Vcal = \Scal \cup \Ccal$.  An edge is present in this network between a shareholder and a company if the shareholder has shares in that company.

In practice our work focusses on a projection of the corporation-shareholder network onto just the shareholder nodes.  That is we define the \paperdefn{Shareholder network} $\Pcal$ to have a set of nodes $\Scal$, the set of shareholders.  An edge between two shareholders, say $s_{i}$ and $s_{j}$, exists in this network if both $s_{i}$ and $s_{j}$ have invested in the same company (at a level above our threshold). In terms of an adjacency matrix $\Pmat$ for this network, we have that
\begin{equation}
  P_{s_{i}s_{j}}
  =
  \begin{cases}
  1 & \mbox{if }  \quad \sum_c B_{s_{i}c} B_{s_{j}c} > 0   \mbox{ and } s_{i} \neq s_{j}\\
  0 & \mbox{if }  \quad \sum_c B_{s_{i}c} B_{s_{j}c} = 0   \mbox{ or }  s_{i} = s_{j}
  \end{cases}
  \, .
  \label{Pmatdef}
\end{equation}
This ensures the shareholder network $\Pcal$ is a simple network.


%
%
%
%
%
%

\subsection{Betweenness Centrality}

A \paperdefn{walk} is a sequence of vertices  in which each node is connected by an edge to the next node in the sequence. A \paperdefn{path} is a walk in which no node appears twice. The \paperdefn{length of the path} is the number of vertices minus one, i.e.\ the number of edges traversed as one moves through the sequence of vertices.

For many centrality measures we consider the shortest path from an initial source node $s$ and ending with a target node $t$. 
The number of shortest paths from $s$ to $t$ is denoted by $\sigma_{st}$ as there can be more than one path of the same length between any pair of vertices. Given these shortest paths, we define $\sigma_{st}(v)$ to be the number of these shortest paths which pass through some $v$ other than $s$ or $t$.
Then, the \paperdefn{betweenness} \cite{freeman1978centrality,newman2010networks} $b(v)$ of a node $v \in \Scal$ is defined to be
\begin{equation}
  b(v) = \sum_{s\neq v\neq t \in V} \frac{\sigma_{st}(v)}{\sigma_{st}}
 \, .
\end{equation}

\subsection{Closeness centrality}

We will define \paperdefn{closeness centrality} $c(v)$ \cite{bavelas1950communication,newman2010networks} of a vertex $v$ to be
\begin{equation}
 c(v) = \frac{n-1}{\sum\limits_{u=1}^{n-1}d(u,v)},
\label{eq:close}
\end{equation}
where $d(u,v)$ is the shortest path distance between $u$ and $v$ and $n$ is the number nodes in the component connected to node $n$.



\subsubsection{Estimating Closeness}\label{closecalc}

Consider first a general random graph, that is, one with a specific degree distribution but otherwise unconstrained, working in large sparse graph regime, $N \rightarrow \infty$, $\texpect{k} \sim O(1)$.  This type of configuration model graph can be constructed using edge rewiring.
Suppose we start at a node of degree $k$. Then we might estimate that the number of nodes $\ell$ steps away from our starting node is
\bea
  n_\ell &=& \zbar^{\ell-1} k \, , \qquad \ell \geq 1 \, ,
  \label{nellformula}
\eea
where $\zbar$ is some effective \paperdefn{branching ratio}. That is we expect each node we arrive at $\ell$ steps away from our starting node, in some breadth first search out from the initial node, to be connected to an average of $\zbar$ new vertices which are then $(\ell+1)$ steps away. The approximation here is that all nodes look the same as they must in a true random graph. The exception is the first node where we know that that has $k$ neighbours if that node has degree $k$. However we note that statistically, all we are really saying in this approximation is that for most networks, taking a few steps is sufficient to allow us to sample any part of the network so statistically many networks will appear to be homogeneous on larger scales.

If we are being more precise, for a random graph near its phase transition, where we can assume a tree like structure, we know that $\zbar$ will be the average degree of a neighbouring node minus one --- we arrive on one edge going into a neighbour, leave on the remaining edges. Because the current degree of a neighbour $\sum_{k}k\frac{kp(k)}{\langle k\rangle} = \frac{\langle k^2 \rangle}{\langle k \rangle}$. So
\bea
 \zbar &=& \frac{\texpect{k^2}}{\texpect{k}} -1\,.
 \label{randomz}
\eea
However, for any given large network, we do not need to assume \eqref{randomz} is true, merely that there is some effective branching ratio such that \eqref{nellformula} still works well.

To estimate closeness, we first estimate the maximum distance $\ellmax$ by demanding that the total number of nodes connected to our starting node is the number in the Largest Connected Component $\NLCC$ as we assume we are studying nodes in this component. This may be estimated as
\bea
\NLCC &\approx&
 \sum_{\ell=0}^{\ell_\mathrm{max}} n_\ell
= 1 + k \frac{(\zbar^{\ell_\mathrm{max}}-1)}{(\zbar-1)}
 \label{Nellest}
\eea
Rearranging for $N_{LCC} \gg 1$, we find that
\bea
\ell_\mathrm{max} (k)
& \approx & \frac{\ln(1+\NLCC(\zbar-1)/k)}{\ln(\zbar)}  \, .
\eea
Not surprisingly, if you start from a high degree node, a high $k$, your first step will reveal far more of the network and so take you closer to the remaining parts.  Thus the maximum distance in a random graph drops as the degree $k$ of the node increases.

Now we can use this to find the closeness $c(v)$ of a node $v$ since this is defined to be the inverse of \paperdefn{farness}, $f(v)$, the average distance from a node to all other nodes.
For the random graphs, or graphs which appear homogeneous on larger scales, we can estimate farness using \eqref{nellformula} as
\bea
 f(v) &=&
 \frac{1}{(\NLCC-1)} \sum_{\ell=1}^{\ell_v} \ell n_\ell
 \approx \frac{1}{\NLCC} k_v \left( \frac{(\ell_v + 1)\zbar^{\ell_v}}{\zbar-1} - \frac{\zbar^{\ell_{v}+1} - 1}{(\zbar-1)^2}\right)
\eea
where we have used \eqref{Nellest} and we write $\ell_v=\ell_\mathrm{max}(k_v)$ as the largest of the shortest path lengths from vertex $v$ which has degree $k_v$.
Not surprisingly this is dominated by the distance to the further nodes as in the tree they are the dominant contribution. We see that if $(\zbar -1) \gg k/N$, i.e.\ if we are not close to the transition and we have a large $N$, then this result for farness gives us that $f(v) \approx \ell(v)$ so that
\bea
  f(v)
 &\approx&
    \frac{\ln(\NLCC(\zbar-1)/k_v)}{\ln\zbar} \, .
\eea
While in this limit a random graph, let alone a real graph, is not a tree, it shows that we should expect the closeness centrality measure to be correlated with the degree of a node.  Indeed the prediction is that the inverse closeness (farness) should show a linear dependence on the logarithm of the degree of a node, $\ln(k)$, with a slope that is the inverse of the log of the branching ration minus one, $1/\ln\zbar$, that is
\beq
 \frac{1}{c(v)} = f(v) = -\frac{1}{\ln\zbar}\ln(k_v) + a \, .
\eeq
Since this expression is true where we do not have a tree, we do not expect the slope to match a the value of $\zbar$ in a random tree \eqref{randomz}.  Rather, if we do find a linear relationship for the farness and logarithm of degree, then the slope is a way of defining an effective branching ratio.

\subsection{Community detection algorithms}\label{appcommdetect}

The Louvain algorithm \cite{BGLL08} aims to produce a community structure which has a large value of modularity $Q$ where
\begin{equation}
   Q = \frac{1}{2m}\sum_{ij}\left[A_{ij} - \frac{k_ik_j}{2m}\right]\delta(c_ic_j)
   \, .
\end{equation}
Here $A_{ij}$ represents the adjacency matrix between nodes $i$ and $j$;
$k_i$ and $k_j$ are the sum of the weights of the edges attached to nodes i and j, respectively;
$m$ is the total number of edges in the graph.
$c_i$ and $c_j$ are the communities of the nodes.
The Louvain algorithm \cite{BGLL08} starts each node in an individual community and tries to increase modularity by moving a node into the community of a neighbour. When a local maximum is reached, the communities are used to define a new graph where each node in the new network represents a single community in the previous network, and the process is repeated.

The Infomap community detection is based on the movements of a random walker. The aim is to choose communities which minimise the amount of information needed to record the movement of random walkers between communities. This is done using the map equation:
\begin{equation}
L(M) = q_\curvearrowleft H(Q) + \sum_{i=1}^m p_{i\circlearrowright}H(P_i),
\end{equation}
where M is the modules or partitions of the network and each node is assigned to a module $i$. $L(M)$ is the description length of the trajectory of a random walker walking along the links of the networks. $q_{i\curvearrowleft}$ and $q_{i\curvearrowright}$ represent that the random walker enters and exits each module i, respectively. For details see Rosvall and  Bergstrom~\cite{rosvall2008maps}.

\subsection{Comparison of community detection results for largest component of Turkey}

If the structure of communities in the data is well established then using two different detection methods should be able to give similar results~\cite{lancichinetti2008benchmark}. After detecting the communities of the graphs using the two algorithms, Louvain \cite{BGLL08} and Infomap \cite{rosvall2008maps}, for two countries, we found that the percentage of nodes whose two communities contain the same nodes is about 75\% in Turkey. However, it is noticed 
that the Louvain method \cite{BGLL08} produces a very large community size, while Infomap does not have this large community. 

\begin{figure}[H]
\centering
\includegraphics[width=\textwidth]{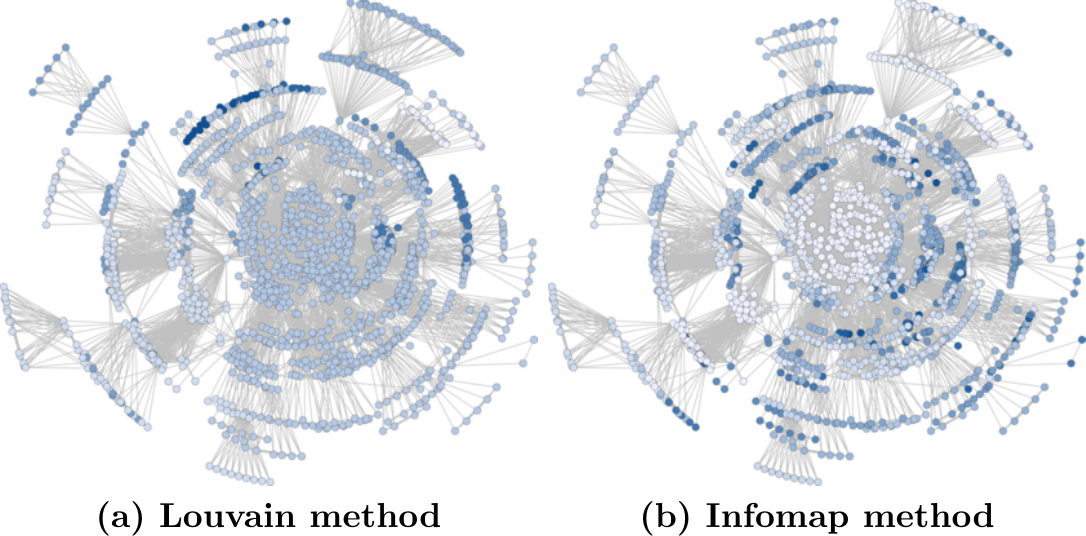}
\caption{{\bf Comparison between the two detection methods.} The left one is for Louvain method and the right one is for Infomap method. The layout style is based on force-directed graph drawing. The number of unique communities for Louvain method is 9 and for Infomap is 124. Each colours represents a community and the colour schemes of the two methods are the same.}
\label{fig:lc}
\end{figure}

If we look at the largest component, the two different methods are separating this component in different ways, see Figure \ref{fig:lc}. We can see from \Figref{fig:lc}, most outside parts of the circles are drawn the same shape of nodes in the same colours which means the these nodes are in one community in both methods. In the center of the graphs, that nodes are coloured differently show these square nodes are in same community in Louvain but in different communities in Infomap method. In Table~\ref{table:com_detail} we give out the statistics of the comparison of communities.
\begin{table}[H]
  \small
  \centering
  \begin{tabular}{|m{1.1cm}|m{2.8cm}|m{1.3cm}|m{1.7cm}|m{1.2cm}|m{1.3cm}|m{2.0cm}|} 
    \hline
    	& & Infomap & Percentage & Louvain & Identical & Ranking of Louvain \\ \hline
    	1st Largest & Size Community & 130 & 100\% & 1199 &  & 1st \\
    	& Types & 9 &  & 15 &  &  \\ \hline
    	2nd Largest & Size Community & 65 & 100\% & 93 &  & 3rd \\
    	& Types & 5 &  & 5 &  &  \\ \hline
    	3rd Largest & Size Community & 58 & 100\% & 58 & yes & 6th \\
    	&  Types & 4 &  & 4 &  &  \\ \hline
    	4th Largest & Size Community & 58 & 100\% & 75 &  & 4th \\
    	& Types & 4 &  & 4 &  &  \\ \hline
    	5th Largest & Size Community & 56 & 100\% & 56 & yes & 7th \\
    	& Types & 5 &  & 5 &  &  \\ \hline
    	6th Largest & Size Community & 51 & 100\% & 1199 &  & 1st \\
    	& Types & 5 &  & 15 &  &  \\ \hline
    	7th Largest & Size Community & 41 & 100\% & 41 & yes & 13th \\
    	& Types & 4 &  & 4 &  &  \\ \hline
    	8th Largest & Size Community & 38 & 100\% & 132 &  & 2nd \\
    	& Types & 4 &  & 6 &  &  \\ \hline
    	9th Largest & Size Community & 38 & 100\% & 38 & yes & 15th \\
    	& Types & 4 &  & 4 &  &  \\ \hline
    	10th Largest & Size Community & 37 & 100\% & 37 & yes & 17th \\
    	& Types& 4 &  & 4 &  &  \\ \hline
  \end{tabular}
  \caption{ {\bf Table for comparison between algorithms.} Table for comparison between the consisting companies in large community for Louvain and Infomap algorithms applied to Turkish network. It is ordered by the community size of the results of Infomap, the percentage 100\% means this Infomap community is the subset of Louvain community in this row. The ranking of Louvain reveals the size ranking of this Louvain community.}
  \label{table:com_detail}
\end{table}

\subsection{Louvain analysis of Individuals and Industrial in Turkey}
\begin{figure}[H]
\centering
\includegraphics[width=\textwidth]{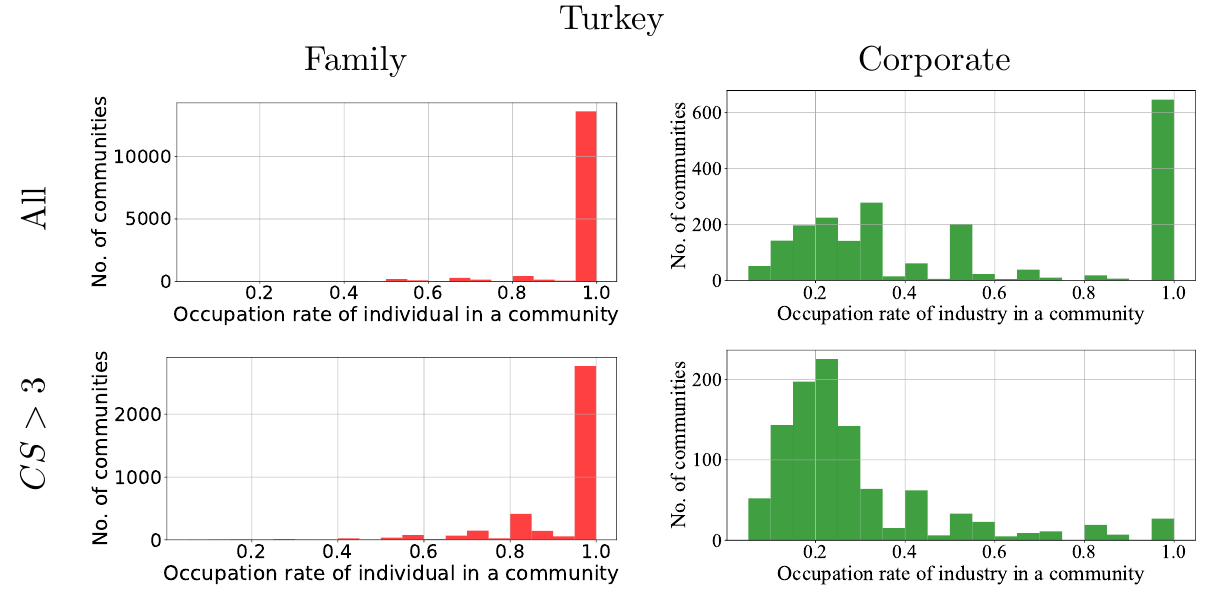}
 \caption{{\bf The bar plots for Louvain community analysis.} The bar plots of frequency analysis for One or more named individuals or families (upper ones), Corporate company (lower ones) in Turkey: Comparison between the frequencies of percentages of this type of owners within one community. The method used is Louvain. The figures in first row analyses all the community sizes while those in second row excludes small communities(including $CS \geq 3$)}
  \label{fig:2d_sec_lou}
\end{figure}



\subsection{Update of Data Base}

The data of two countries is retrieved from BvD~\cite{BvD}, which is updated every year. The total number of known companies in a given year changes. For example, a 4\% difference is observed from 2017 to 2018. However, the authors have downloaded the data and done the analysis at different years from 2016, 2017 and 2018. The results described in the main text show no noticeable differences.

\clearpage

\end{document}